\documentclass[manuscript]{aastex62}

\usepackage{graphics,graphicx}
\usepackage[caption=false,font=footnotesize]{subfig}
\usepackage{txfonts}
\usepackage{textcomp}
\usepackage{epstopdf}
\usepackage{lineno}
\usepackage{arydshln}
\graphicspath{{./}}

\submitjournal{ApJL}

\shorttitle{SEP pivot energy for Mars' surface radiation}
\shortauthors{Guo et al.}

\begin{document}

\title{The pivot energy of Solar Energetic Particles Affecting the Martian surface radiation environment}

\correspondingauthor{Jingnan Guo}
\email{jnguo@ustc.edu.cn}

\author{Jingnan Guo}
\affiliation{School of Earth and Space Sciences, University of Science and Technology of China, Hefei, PR China}
\affiliation{CAS Center for Excellence in Comparative Planetology, Hefei, PR China}
\affiliation{Institute of Experimental and Applied Physics, Christian-Albrechts-University, Kiel, Germany}

\author{Robert F. Wimmer-Schweingruber}
\affiliation{Institute of Experimental and Applied Physics, Christian-Albrechts-University, Kiel, Germany}

\author{Yuming Wang}
\affiliation{School of Earth and Space Sciences, University of Science and Technology of China, Hefei, PR China}
\affiliation{CAS Center for Excellence in Comparative Planetology, Hefei, PR China}

\author{Manuel Grande} 
\affiliation{Unversity of Aberystwyth, Aberystwyth, UK}

\author{Daniel Matthi\"{a}}
\affiliation{German Aerospace Agency, Cologne, Germany}

\author{Cary Zeitlin}
\affiliation{Leidos Innovations Corporation, Houston, TX, USA}

\author{Bent Ehresmann}
\affiliation{Planetary Science Division, Southwest Research Institute, Boulder, CO, USA} 

\author{Donald M. Hassler}
\affiliation{Planetary Science Division, Southwest Research Institute, Boulder, CO, USA} 

\begin{abstract}
Space radiation is a major risk for humans, especially on long-duration missions to outer space, e.g., a manned mission to Mars. Galactic cosmic rays (GCR) contribute a predictable radiation background, the main risk is due to the highly variable and currently unpredictable flux of solar energetic particles (SEPs). 
Such sporadic SEP events may induce acute health effects and are thus considered a critical mission risk for future human exploration of Mars.
Therefore, it is of utmost importance to study, model, and predict the surface radiation environment during such events. It is well known that the deep-space SEP differential energy spectrum at high energies is often given by a power law.
We use a measurement-validated particle transport code to show that, for large SEP events with proton energy extending above $\sim$ 500 MeV with a power-law distribution, it is sufficient to measure the SEP flux at a pivot energy of $\sim$ 300 MeV above the Martian atmosphere to predict the dose rate on the Martian surface. In conjunction with a validation by in-situ measurements from the Martian surface, this remarkable simplification and elegant quantification could enable instant predictions of the radiation environment on the surface of Mars upon the onset of large SEP events.
\end{abstract}


\section{Introduction} \label{sec:intro}
Galactic Cosmic Rays (GCRs) and Solar Energetic Particles (SEPs) are the two major types of energetic particles in the heliosphere that may impose radiation risks to humans on space and planetary missions. 
On the surface of Earth, we are mostly protected against such high-energy particles from space thanks to Earth's magnetosphere and our sufficiently thick atmosphere which can deflect and stop the large majority of the high energy charged particles.
However, the surface of Mars is much more exposed to highly energetic particles than Earth because of the lack of a global magnetic field and its very thin atmosphere. With significantly less  shielding, exposure to the radiation environment on the surface of Mars remains a major concern and health risk for future human explorers \citep[e.g.,][]{cucinotta2006, hassler2014}.

GCRs are mainly composed of protons and helium ions \citep{simpson1983} and are omnipresent as they arrive in the solar system from interstellar space. They are modulated by heliospheric magnetic fields which evolve dynamically as solar activity varies in time and space, with a well-known 11-year cycle \citep[e.g.,][]{parker1958interaction}. 
In contrast to GCRs, SEPs -- mainly protons and electrons -- are emitted from the Sun and accelerated by sporadic solar eruptions such as flares and/or Coronal Mass Ejection (CME) associated shocks \citep[e.g.,][]{lario2005}. 
SEP events are often impulsive and could enhance the radiation level significantly especially at places which are magnetically connected to the particle injection site and where sufficient shielding against radiation is unavailable, such as on the Martian surface \citep[e.g.,][]{guo2018generalized}. 
As for near-future human exploration missions to Mars, the immediate forecast (or nowcast) of the radiation environment on the surface of Mars during periods of enhanced solar activity is essential and critical. 

Hitherto, SEP measurements on the surface Mars are very scarce and time-limited.
The Mars Science Laboratory (MSL) rover Curiosity landed on Mars on August 6, 2012 \citep{grotzinger2012mars}. Since then, its Radiation Assessment Detector \citep[RAD,][]{hassler2012} has been measuring the Mars surface radiation environment which so far has been dominated by the GCR component. 
To date only 6 SEP events have been detected on Mars, and of these the September 10 2017 solar eruption associated SEP event is the most intense event seen by RAD \citep{ehresmann2018energetic, zeitlin2018, guo2018modeling, hassler2018} and is also the first and only solar particle event that caused ground-level enhancement detected at two different planets, Earth and Mars \citep{guo2018modeling}. 
Nevertheless, this event is not intense enough to pose any radiation risk to the health of an astronaut at the ground level of Mars \citep{zeitlin2018}.

This sparsity of surface data for SEP-induced radiation requires us to resort to sophisticated physics-based modeling thereof. 
In some previous studies \citep{matthia2016martian, MATTHIA201718, guo2019atris} and as also described in Appendix A, we developed a GEANT4 particle transport model implemented with Martian atmospheric and regolith properties and benchmarked this model using RAD measurements.  Here we use this validated model to calculate the surface radiation for different input spectra at the top of the atmosphere.  
In particular, we calculate the induced Martian surface radiation by a variety of SEP events with different properties such as their energy range, intensity, power-law index, etc.
For the first time, we find a pivot energy ($\sim$ 300 MeV) at which the SEP flux alone can be used to determine the Martian surface dose rate. I.e., with a fixed flux at this pivot energy, the variation of the power-law spectral index does not affect the surface radiation. 
This finding advances our understanding of the radiation risks during possibly-adverse space weather conditions. Together with SEP injection and inter-planetary transport models, we can provide instantaneous and quantitative alerts for future human missions at Mars upon the onset of large SEP events at the Sun.

\section{Method}\label{sec:method}
\begin{figure}[htb!]
	\centering
	\begin{tabular}{cc}
		\subfloat{\includegraphics[trim=0 0 0 0,clip, scale=0.4]{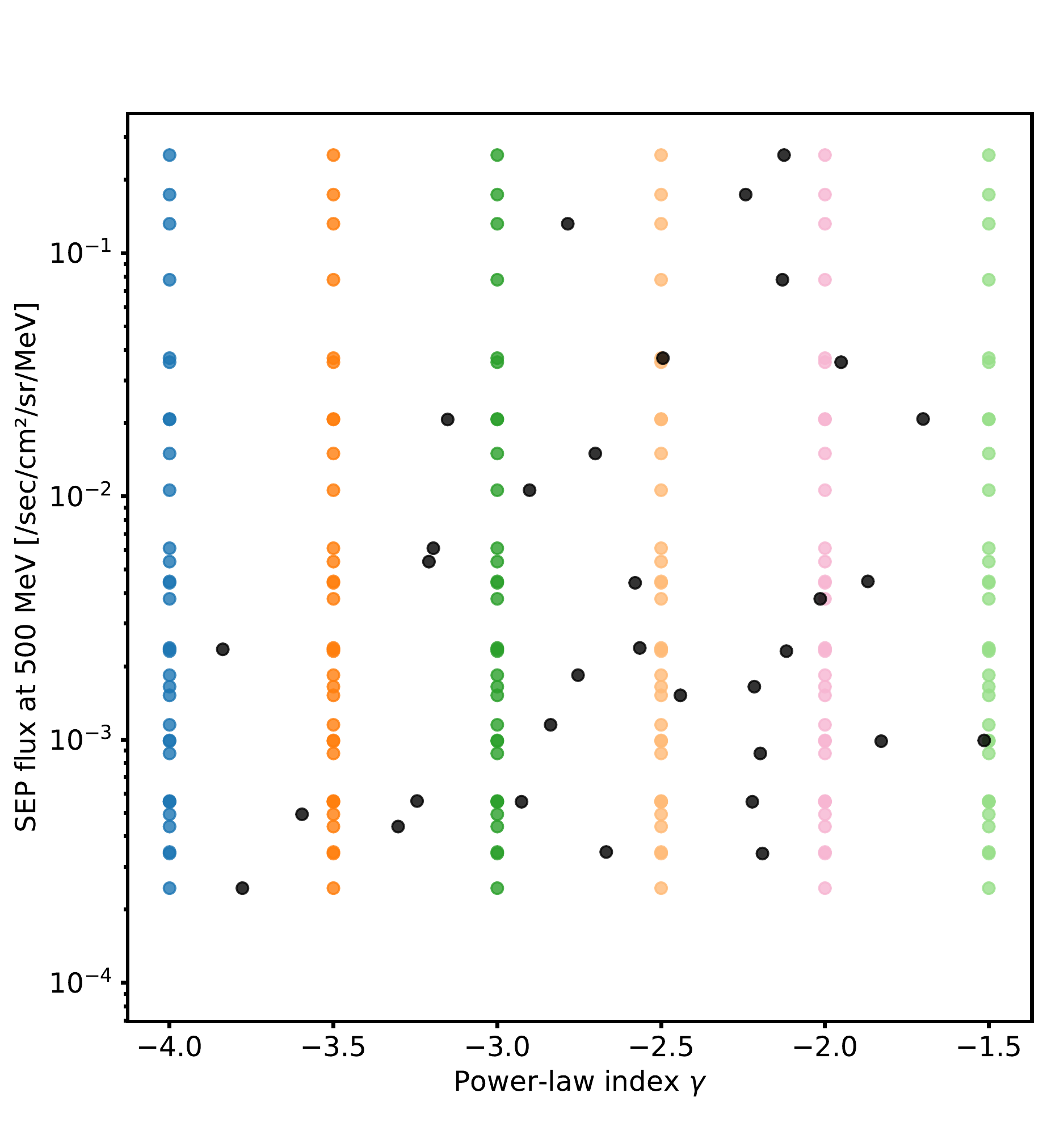} }& 		
		\subfloat{\includegraphics[trim=0 0 0 0,clip, scale=0.4]{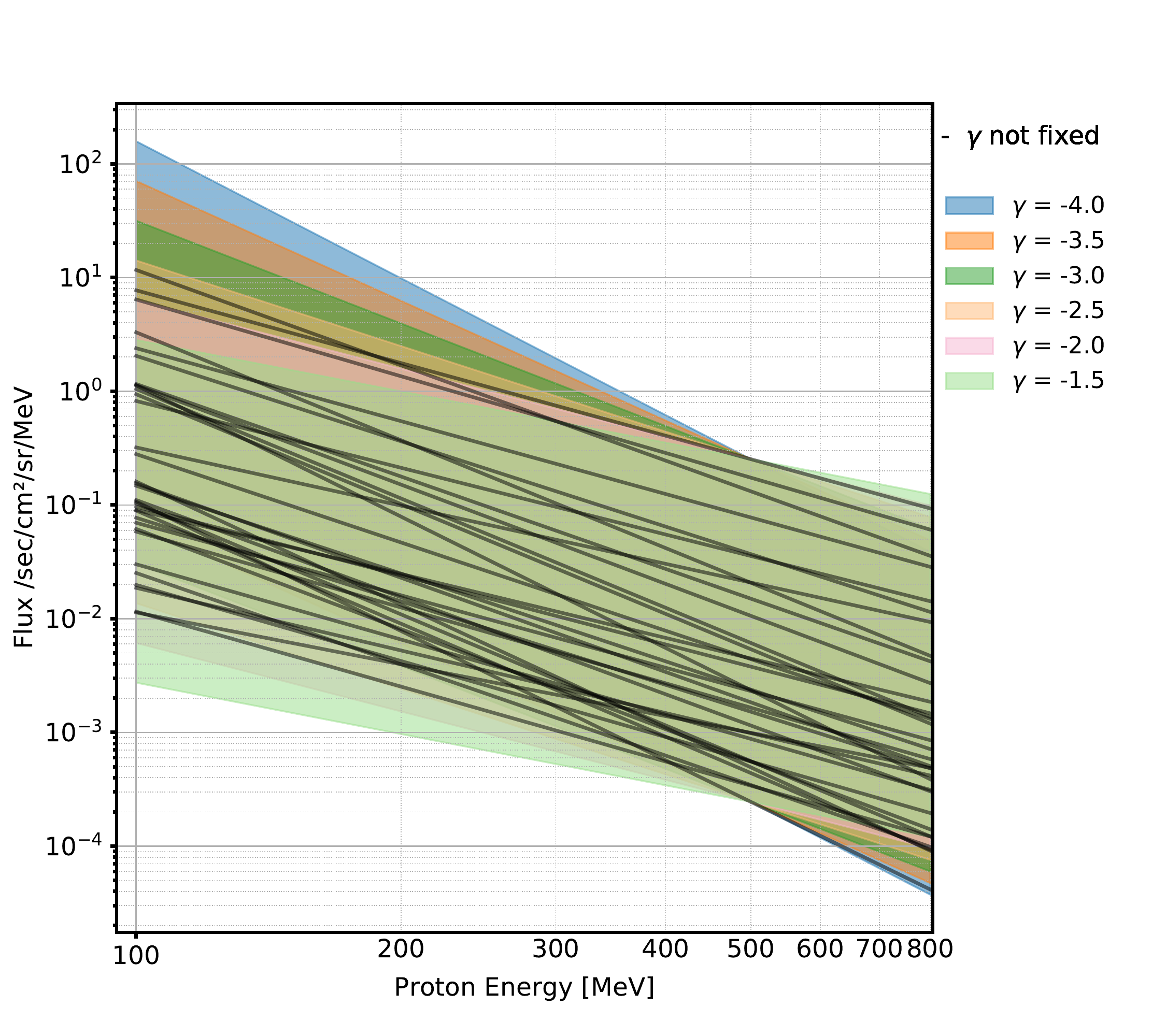}} \\
	\end{tabular}
	\caption{Left: SEP spectral intensity at 500 MeV versus spectral index $\gamma$ of SEP events used in this study. Right: SEP spectra in the range of 100-800 MeV. Black points (left panel) or black lines (right panel) are the power-law spectra of the original 33 events from \citet{kuehl2016}. Different coloured ranges are the regenerated spectra with each colour representing a set of SEP spectra with the same power-law index $\gamma$, but different intensities. }\label{fig:SEP_powerlaw}
\end{figure}
In this study, we calculate the induced Martian surface radiation by a variety of SEP events with different properties such as their energy range, intensity, power-law index, etc.
Then we analyze and correlate the parametrized properties of the SEP spectra and the resulting surface dose rate of each event in order to find a simplified quantification of the Martian surface radiation based on SEP spectral properties.
Starting from a list of more than 30 significant SEP events detected in situ at Earth by SOHO in 20 years \citep{kuehl2016}, each event spectrum from 100 to 800 MeV was derived in a 2hr interval starting 30 minutes after the event onset (in an energy channel covering 100 MeV to 1 GeV) and then fitted with a power-law distribution of $I(E) = I_0 E^\gamma$, shown as black lines in Fig. \ref{fig:SEP_powerlaw}. The fitted power-law parameters of each event and the goodness of the fit ($R^2$ mostly larger than 0.9) are listed in Table 3 of \citet{kuehl2016}.
It is not certain that these events reached Mars, and even if they did, the SEP spectra were probably different from those measured at Earth.
This is because the observed SEP spectra and intensity depend on the magnetic connections of the planets/spacecraft to the acceleration/injection locations and also the propagation of SEP particles in the interplanetary space.
However,  for the purpose of a statistical and parametric study for how different SEP properties may affect the Martian radiation environment, we use these near-Earth spectra at Mars to obtain a better understanding of SEP-induced radiation on Mars.

Based on each fitted SEP spectrum, we calculate its induced secondary spectra and accumulated radiation dose rate on the surface of Mars in a 0.5 mm water slab (an approximation of thin skin structures of a human body).
In order to quantify the dependence of the surface radiation on SEP properties, we also use SEP spectra with the same spectral power-law index, but with different intensities.
Due to the lack of statistics of listed events with the same power-law index, we "re-generate" a set of events by fixing the intensity of each SEP spectra at 500 MeV, while forcing all spectra to have the same power-law index, e.g., $\gamma=-2$. The choice of the anchor point 500 MeV is arbitrary, as this is just helping to regenerate the spectra with similar intensities;  it could also be 200 or 300 MeV.
The regenerated spectral indices and the particle energy spectra are shown in Fig. \ref{fig:SEP_powerlaw}.

Apart from the SEP intensity and spectral index, we also investigate the influence of the energy ranges on the surface radiation. 
In \citet{guo2018generalized}, for a few historical event, we calculated the surface dose rate resulting from 100-800 MeV primary protons is about 93\% of that induced by the full SEP spectra. 
This is because of the effective atmospheric shielding of particles with low energies \footnote{As shown in Fig.2 in \citet{guo2019readyfunctions}, there is a sharp decrease of surface dose contribution from primary protons with energies below $\sim$ 165 MeV $\pm$ 30 MeV (depending on the elevation of interest and the varying seasonal pressure), which can be considered to be the atmospheric cutoff energy.} and the nature of the SEP spectra where the high energy component has significantly less flux.
In this study, we first calculate the Martian surface radiation induced by SEP spectra from 100 MeV to 800 MeV as used for fitting the SEP events by \citet{kuehl2016}. 
We then test the influence of the SEP energy range on the surface radiation using three different cases: 15-1000 MeV, 15-1500 MeV and 15-2000 MeV.
Based on these spectra with different properties, we model the induced Martian surface radiation, dose rate and dose equivalent rate by each event as shown in Section \ref{sec:results}. 

\section{Results and Discussions}\label{sec:results}
\begin{figure}[htb!]
	\centering
	\begin{tabular}{cc}
		\subfloat{ \includegraphics[trim=7 15 60 35,clip, scale=0.43]{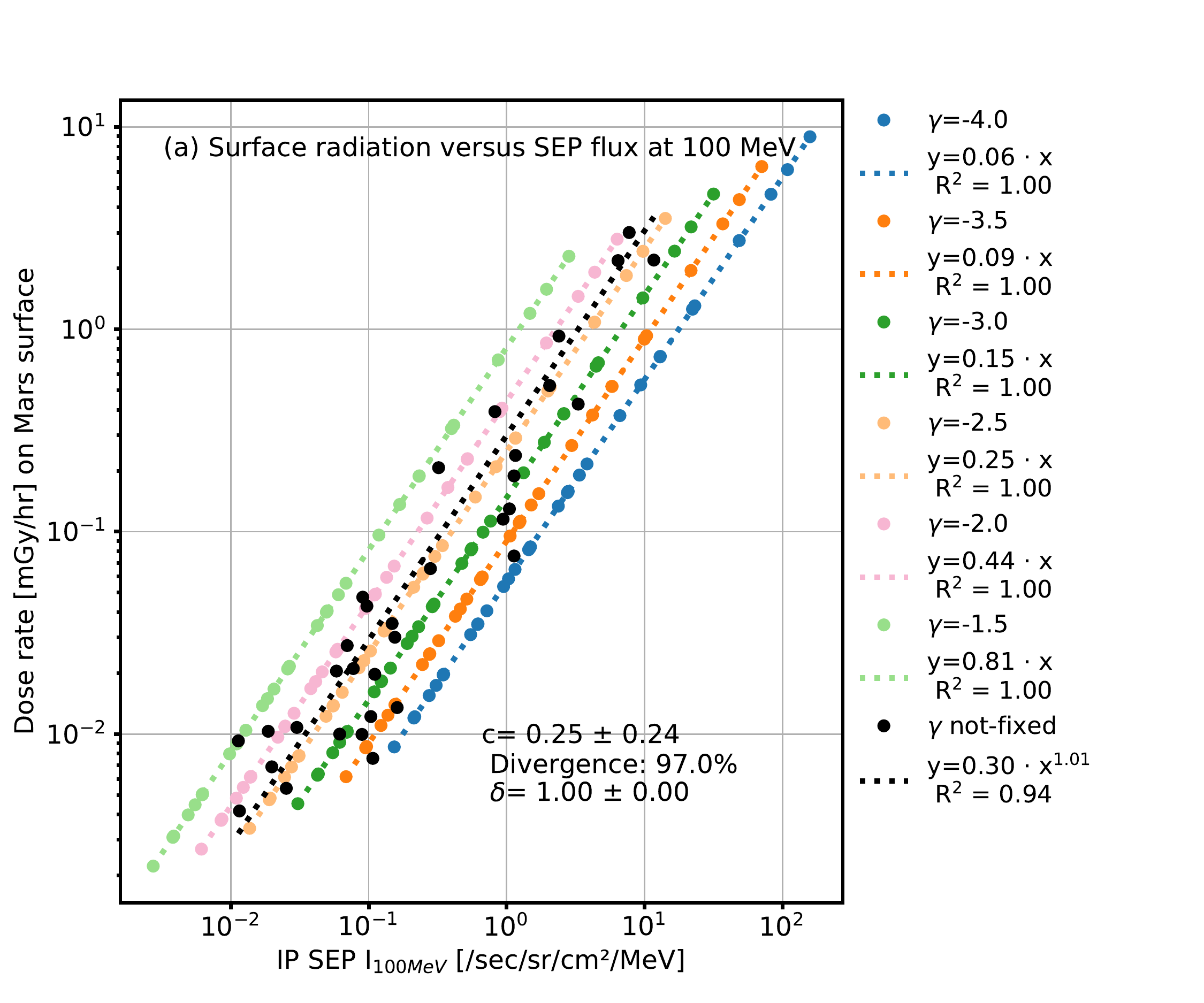} } & 		
		\subfloat{ \includegraphics[trim=7 15 60 35, clip, scale=0.43]{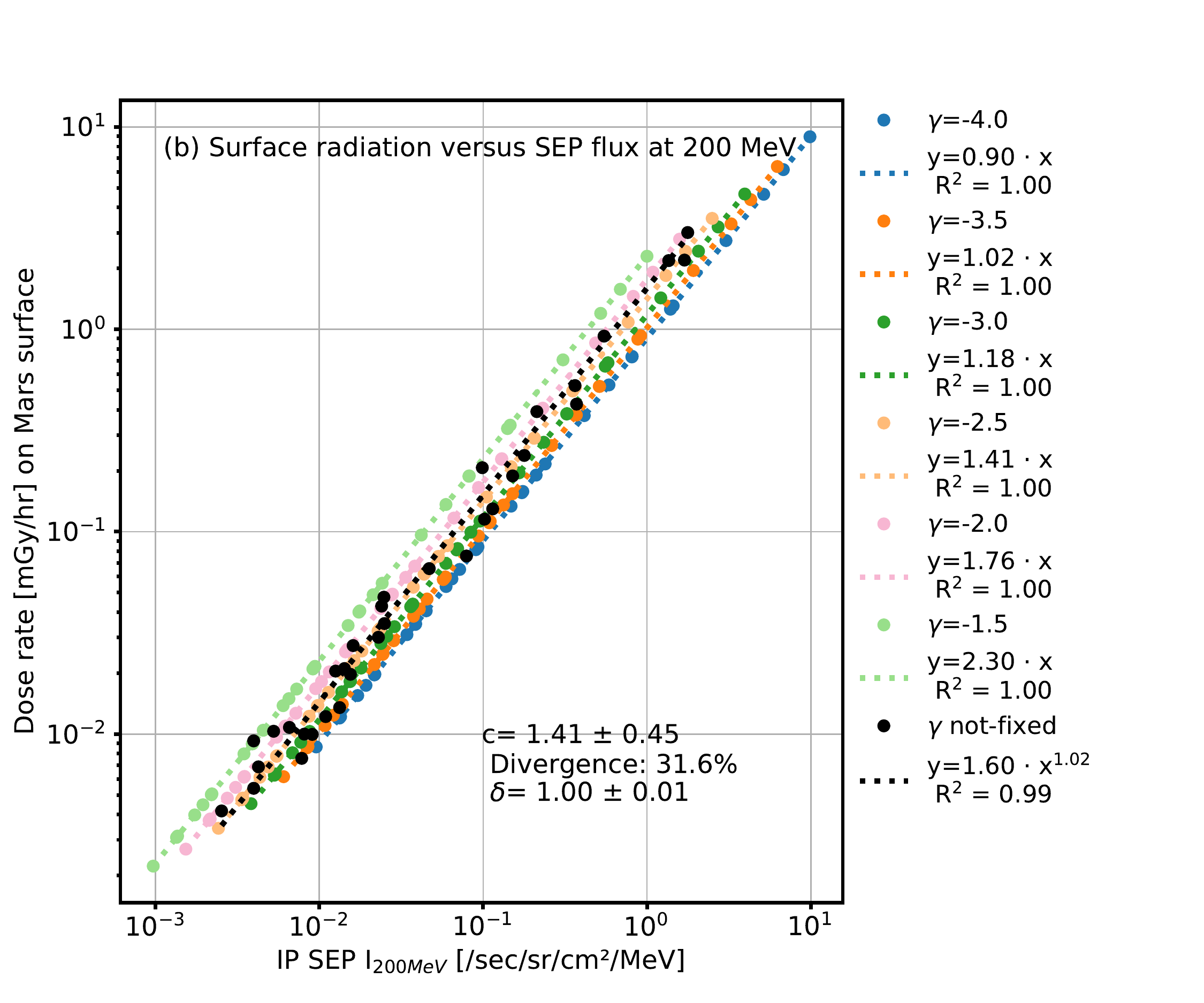} } \\
		\subfloat{ \includegraphics[trim=7 15 60 35,clip, scale=0.43]{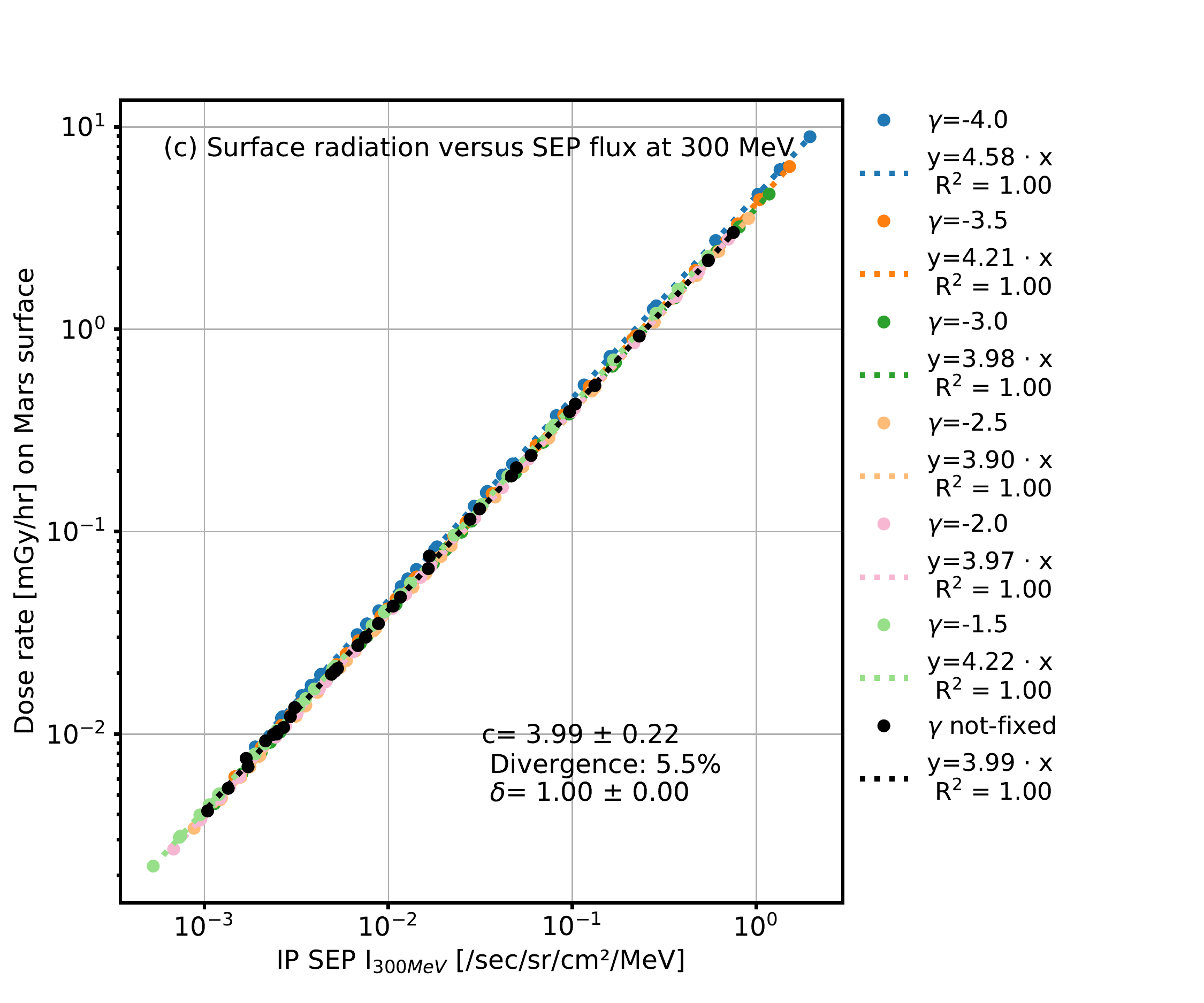} } & 
		\subfloat{ \includegraphics[trim=7 15 60 35,clip, scale=0.43]{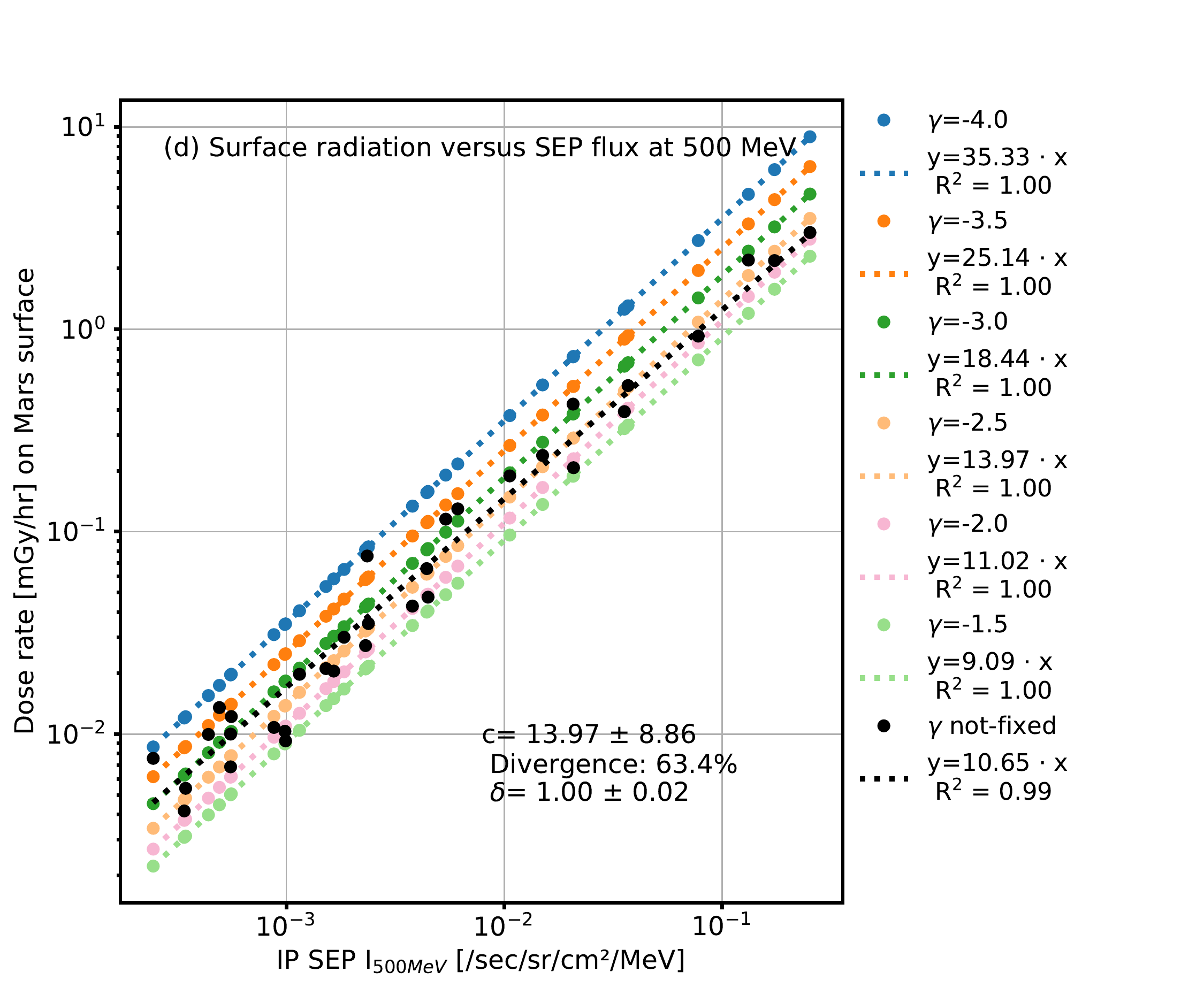} } 
	\end{tabular}
	\caption{Martian surface dose rate resulting from primary SEP spectra (100- 800 MeV protons) versus primary SEP flux at different energies on top of the Martian atmosphere. Black dots are the calculations based on actual events from the list of \citet{kuehl2016} and each different colour represents a fixed power-law index $\gamma$. Data points of the same colour in each panel are fitted with $y=cx^\delta$ with the fitted parameters and their deviations shown in the legends.} \label{fig:pivot_point}
\end{figure}

Fig. \ref{fig:pivot_point} shows the SEP-induced Martian surface dose rate (from primary 100-800 MeV protons) versus the primary SEP flux at certain energies (e.g, 100, 200, 300 and 500 MeV in each panel). 
As shown, when the power-law spectral index is the same for different events, e.g., $\gamma = -4$, the induced Martian surface dose rate and the SEP flux at certain energy are marked as blue dots and fitted with blue dashed lines in each panel of the figure. 
Due to the large span of the x and y values in Fig. \ref{fig:pivot_point}, we plot and fit the results in logarithmic scale $y=cx^\delta$. However as indicated by the fitted index $\delta \sim 1$, the correlation is very close to a perfect linear function. 
For different power-law $\gamma$ indices, the linear correlations between the surface dose rate and the SEP flux at a certain energy are generally different.
For instance, at 100 MeV, as shown in panel (a), the fitted linear coefficient $c$ is different from one another and the divergence (ratio of the standard deviation and the mean value) is as large as 97\%.
This is expected: a power-law SEP spectrum depends on two parameters and when the $\gamma$ value is fixed, the induced dose rate should be proportional to the SEP intensity (i.e., $\delta=1$); when spectral $\gamma$ values are different, the dependence of the dose rate on the intensity, $c$, also changes. 
Surprisingly, at 300 MeV, the divergence dramatically decreases to only 5.5\% (shown as a converging line of all different colours in panel (c)), meaning that the correlation between the surface dose rate and the SEP flux is almost independent on the SEP spectral index $\gamma$. 
In other words, for an SEP spectrum which can be fitted with a power-law distribution between 100-800 MeV, its intensity at 300 MeV alone can determine the Martian surface radiation dose rate.  
 
\begin{figure}[htb!]
	\centering
	\begin{tabular}{cc}
		\subfloat{ \includegraphics[trim=7 15 65 35,clip, scale=0.43]{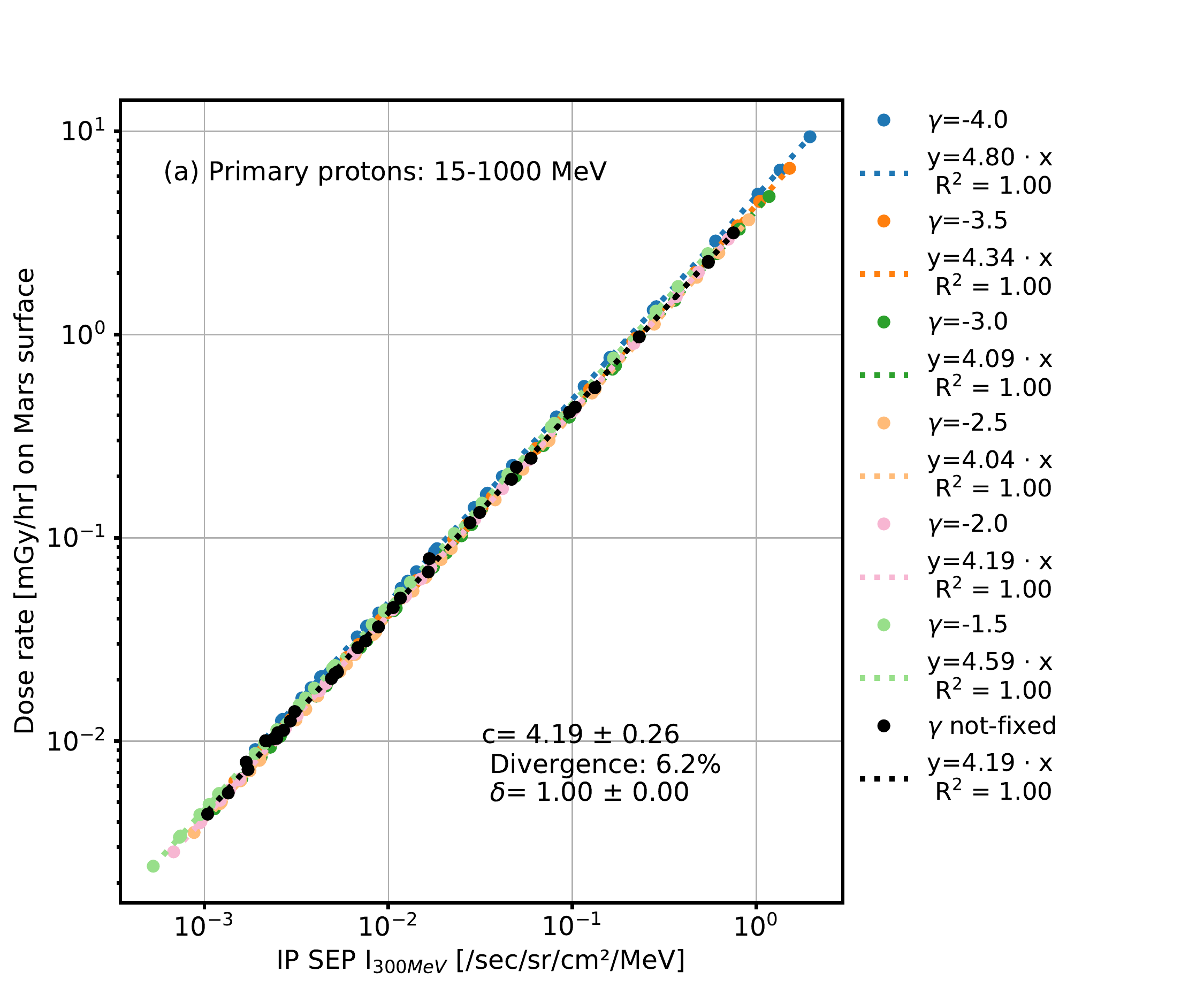} } & 		
		\subfloat{ \includegraphics[trim=7 15 65 35,clip, scale=0.43]{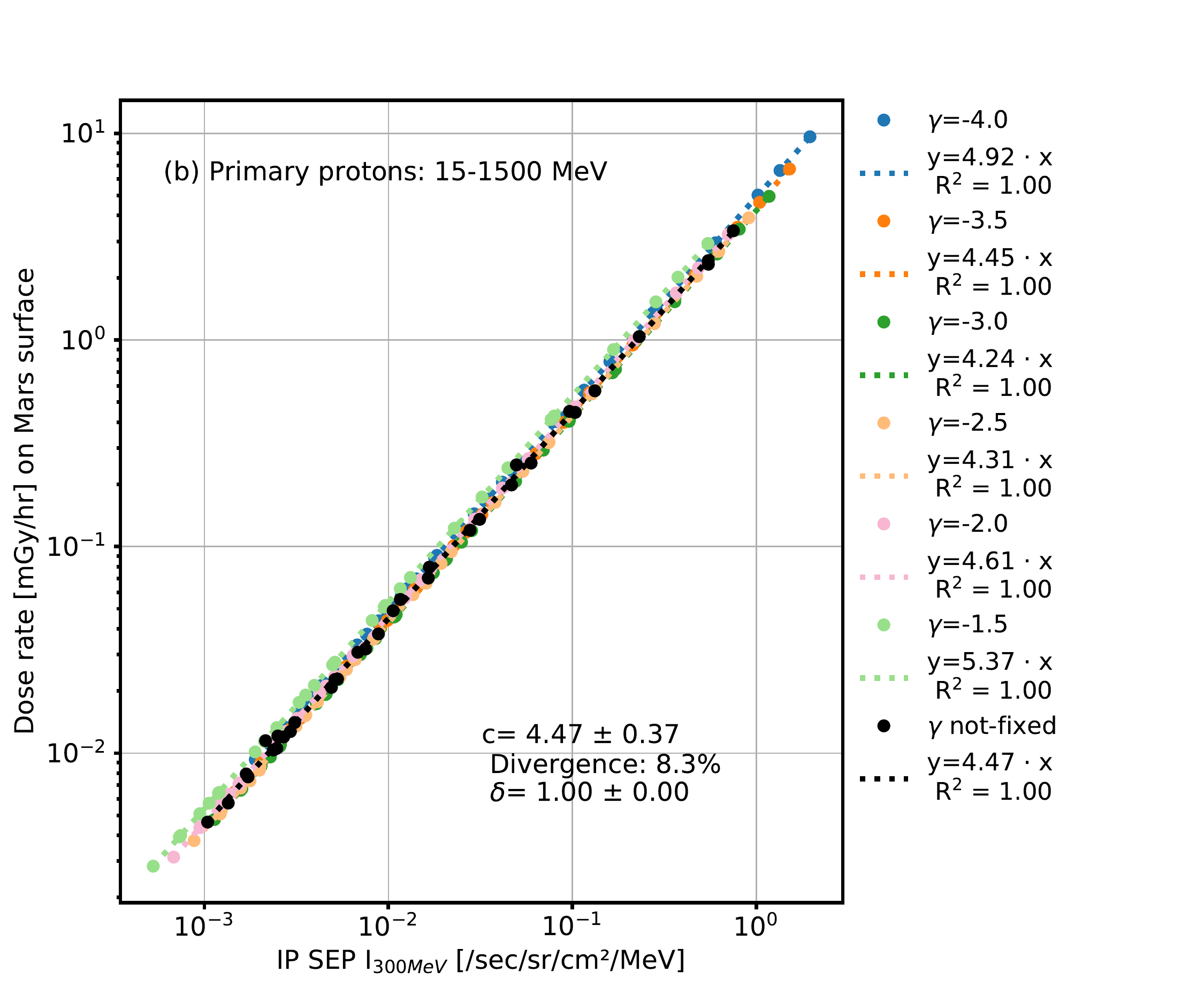} } 
	\end{tabular}
	\caption{Martian surface dose rate versus primary SEP flux at 300 MeV for primary power-law spectra ranging from (a) 15- 1000 MeV protons and (b) 15- 1500 MeV protons. Color features of the figure are the same as in Fig. \ref{fig:pivot_point}.} \label{fig:pivot_point1}
\end{figure}

To test the robustness of this result, we also carried out the above study using other ranges of the incident SEP spectra, i.e., 15-1000 MeV, 15-1500 MeV and also 15-2000 MeV to account for the remaining contribution to the surface radiation by particles below 100 MeV (which can still slightly induce the surface dose via generating some secondary particles arriving at the surface, especially neutrons and gammas \citep{guo2019readyfunctions}) or above 800 MeV when such particles are present during extreme events, especially with flat power-law spectra (small $|\gamma|$ values).
The correlation of the 300 MeV SEP intensity and the Martian surface dose rate is still very good for the three test cases (and Fig. \ref{fig:pivot_point1} shows the results of the first two cases) and the fitted correlation coefficients are rather consistent with that of the 100-800 MeV range shown in Fig. \ref{fig:pivot_point}. 
Considering all four cases of energy ranges, we formulate the correlation between SEP intensity [particles/sec/sr/cm$^2$/MeV] at 300 MeV and the Martian surface radiation dose rate [mGy/hour] as following:
\begin{eqnarray}\label{eq:pivot_dose}
D_{Mars} = 4.45 \cdot  I_{300 MeV} ~~(\pm 9.8\%),
\end{eqnarray}
where the coefficient 4.45 has the unit of (mGy/hr)$\cdot$(sec$\cdot$sr$\cdot$cm$^2$$\cdot$MeV) and is the mean value of the fitted parameters from all cases calculated (4 different energy ranges, each with 7 different power-law indices). The 9.8\% error bar is accounting for the divergence of fitted parameters from different cases. 
We do note that the distribution of all coefficients is slightly different from Gaussian and the median value is 4.33, smaller than the mean value. The upper and lower boundaries of the coefficients from all cases are 5.87 and 3.90, respectively. 

In terms of biological effectiveness associated with radiation exposures on human beings, the dose equivalent is often more referred to for evaluating the deep space exploration risks \citep{icrp60}. 
Therefore we have also explored the correlation between SEP properties, varying among different energy ranges, intensities and power-law indices, and the dose equivalent and find the 300 MeV SEP flux [particles/sec/sr/cm$^2$/MeV] still being the pivot point for the surface dose equivalent rate [mSv/hour] following this correlation:  
\begin{eqnarray}\label{eq:pivot_dose-equ}
E_{Mars} = 5.56 \cdot  I_{300 MeV}  ~~(\pm 12.9\%),
\end{eqnarray}
where the coefficient 5.56 is the mean value of all cases studied and has the unit of (mSv/hr)$\cdot$(sec$\cdot$sr$\cdot$cm$^2$$\cdot$MeV). 

The error bars in Eqs. \ref{eq:pivot_dose} and \ref{eq:pivot_dose-equ} are no bigger than the uncertainties, which are between 5\% and 15\%, resulting from the choice of different atmospheric properties, particle transport models through the Martian atmosphere and physics lists included therein as discussed in \citet{matthia2016martian} and \citet{guo2019atris}. 
Due to this strong linear correlation between the SEP flux at 300 MeV and the Martian surface dose rate and dose equivalent rate, we name the 300 MeV as the pivot energy of SEPs contributing to the Martian surface radiation.
This can be explained by the balance between the contribution to the surface dose from particles below 300 MeV and those above. For instance, when changing the spectral index $\gamma$ from -4 to -2, the reduction of the dose contribution from particles below 300 MeV is compensated by the increased dose from particles above this pivot point.
Alternatively, let's consider the primary proton contribution to the Martian surface dose as a function of the incoming particle energy which has been calculated and shown in Fig. 2 of \citet{guo2019readyfunctions}. The function shows that primary protons below $\sim$ 160 MeV have little effect on the Martian surface radiation. 
For an SEP event, we need to fold its spectra with this energy-dependent function for calculating the surface dose rate. 
The combination of an SEP spectrum, which often has a power-law distribution at energies above $\sim$ 10-30 MeV, with the Martian atmospheric function leads to an energy point at which the dose contributions below and above this point are balanced out no matter how the spectral index changes. In Appendix B we show the analytical derivation of the pivot energy of the dose rate induced by power-law SEP spectra. 

We note that this result is limited to power-law shaped large SEPs with proton energy extending to above $\sim$ 500 MeV as a common feature of the sampled events for studying the correlation \citep{kuehl2016}.
This empirical correlation will likely not work for SEP events with only protons of energies smaller than $\sim$ 500 MeV {or for events with distributions far away from a power-law shape at these high energies.
We have investigated the influence of non power-law spectra on the correlation using a few historical events: the October 1989, September 1989 and August 1972 events, for which the proton distribution have been constructed using a Weibull function \citep{townsend2006carrington}.
For each of these three events, we calculate the Martian surface dose rate induced by the whole input spectrum and compare it with that estimated by our pivot-energy correlation shown in Eq. \ref{eq:pivot_dose}.
For the two 1989 events, although the SEP spectra do not have a single power law through all energies, particles distributed between $<\sim$ 100 MeV and 1 GeV can be reasonably fitted by a power law with $\gamma$ indices around -2.2 \citep{guo2018generalized}. 
The surface dose rates estimated using the pivot-energy correlation are about 9\% and 6\% larger than those calculated from the complete spectra for the October and September 1989 events, respectively. The overestimation is because the actual spectra, different from a single power law, fall off slightly at both low- ($<\sim$ 30 MeV) and high-energy ($>\sim$ 1 GeV ) ends. However the pivot-energy estimation is still consistent with the actual surface radiation within uncertainties.
Alternatively, the August 1972 spectral shape is rather unusual with a large initial flux at energies below $\sim$ 100 MeV (almost a flat power law with $\gamma$ between 0 and -1) followed by a very sharp fall-off at higher energies; particle flux between 100 and 800 MeV can not be described by any power-law distribution while those above 500 MeV up to 1 GeV can be fitted by an exceptionally soft power law with an index of -38. 
The surface radiation estimated using the 300-MeV pivot-energy correlation is about 17 times smaller than that resulted from the complete spectrum, which is explainable by the characteristics of the spectra where majority of the event flux is contributed by particles below $\sim$ 200 MeV. 
This highlights the importance and necessity of continuing Martian surface radiation measurement by MSL/RAD for nowcasting any potentially hazardous radiation enhancement due to the non-predictable nature of SEPs. }

Another additional caveat is that the Martian atmospheric depth used in the current model is equivalent to the average value of $\sim$ 22 g/cm$^2$ measured at Gale Crater by the Curiosity rover. This depth was chosen for the purpose of model validation (more in Appendix A) using MSL/RAD. The atmospheric shielding may be slightly smaller at other high altitude places for which the correlation obtained here needs to be re-evaluated. 
{Thus further benchmark of the model using in-situ RAD measurements at different atmospheric depth, especially as Curiosity is climbing up to Mount Sharp, would be needed.  
Long-term measurements of the radiation enviroment on Mars under different solar modulation and atmospheric conditions would also favour the validation of various radiation transport models which may differ by up to an order of magnitude \citep{MATTHIA201718}.}

Besides, the dose and dose equivalent given here are based on a "biological phantom" representing some thin skin structures and can not be directly applied to phantoms such as detectors with different materials or body organs embedded under different body depths. 
Nevertheless, such a simplified and elegant quantification can serve to make instant predictions, within one millisecond, of the radiation environment on Mars upon the onset of large SEP events where protons are accelerated up to more than 500 MeV. 

For the purpose of mitigating radiation risks for future Mars missions, a particle detector sensitive to protons at these specific energies (i.e., at 300 MeV and perhaps also a channel for $E> 500$ MeV to detect if particles reached the high energy range) can be located in the inner heliosphere with a good magnetic connection to Mars in order to provide an alert ($\sim$ hours ahead) for potential hazardous radiation environment on Mars. 
However, the 300~MeV particle flux at Mars (needed as our model input) will be likely smaller than that at Earth orbit due to the propagation and scattering of particles as they are transported outwards in the heliosphere. 
Thus, we need detailed knowledge of the acceleration and injection profiles of the particles \citep[e.g.,][]{lario2017solar} as well as the heliospheric structure in the interplanetary space which may influence the particle propagation towards Mars \citep[e.g.,][]{guo2018modeling} to achieve more precise forecasting. 
Such knowledge, especially for the Mars direction, is often not available and thus difficult to be properly included in a timely-forecast model. 
However, in light of much more detailed information on solar eruptions and particle propagations provided by Parker Solar Probe \citep{fox2016psp} and Solar Orbiter \citep{mueller2013solar} in the near future, more reliable data-driven SEP injection and inter-planetary transport models could be used to make better predictions of the SEP flux at Mars, allowing us to provide instantaneous and quantitative alerts for future human missions at Mars upon the onset of large SEP events at the Sun.

\acknowledgments
RAD is supported by NASA (HEOMD) under Jet Propulsion Laboratory (JPL) subcontract 1273039 to Southwest Research Institute and in Germany by the German Aerospace Center (DLR) and DLR’s Space Administration grants 50QM0501, 50QM1201 and 50QM1701 to the Christian-Albrechts-University, Kiel.
J.G. and Y.W. are supported by the Key Research Program of the Chinese Academy of Sciences (Grant No.XDPB11 and QYZDB-SSW-DQC015) and the NSFC (Grant No.41842037).
J.G., R.F.W.S. and M.G. acknowledge the International Space Science Institute, which made part of the collaborations in this paper through the ISSI International Team 353 “Radiation Interactions at Planetary Bodies”.

\bibliography{msl_rad_guo} 

\begin{appendix}

\textbf{Appendix A: Model Implementation for calculating the Martian surface radiation environment}
\\

\begin{figure}[htb!]
	\centering
	\includegraphics[trim=0 0 0 0,clip, scale=0.72]{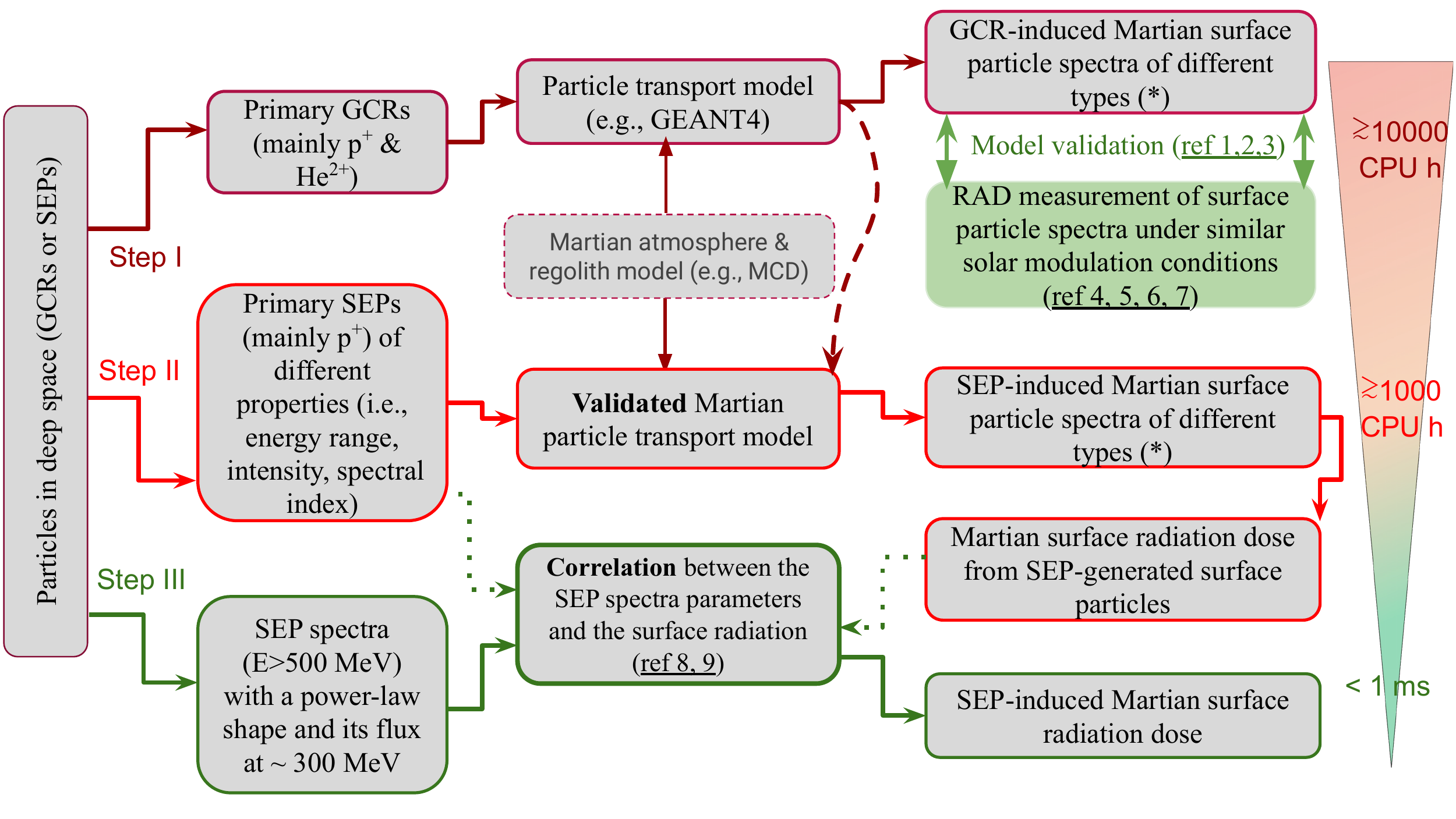} 
	\caption{Flowchart of the implementation of the models for calculating the Martian surface radiation environment as used in the current and relevant studies. Step I includes the model set up and validation. Step II applies the above validated model to alarmed SEP events with different properties. Step III employs the parametrized correlation between SEP spectral property and the surface radiation environment based on Step II analysis to quickly derive the Martian surface radiation based on minimized SEP spectral input. (*) Surface secondary particles considered in the model output include proton, electron, position, $^4$He, $^3$He, deuteron, triton, neutron and gamma above 1 MeV.  References 1, 2, 3 for model validations are \citet{matthia2016martian, MATTHIA201718, guo2019atris}. References 4, 5, 6, 7 for RAD surface spectra measurements are \citet{ehresmann2014, EHRESMANN20173, koehler2014, guo2017neutron}. References 8 and 9 for extracted correlations are \citet{guo2018generalized} and this study. The approximate computing central processing unit (CPU) time is marked in the rightmost part. More explanations and descriptions of this chart can be found in the text.}\label{fig:model_chart}
\end{figure}

For calculating the particle flux and radiation level on the surface of Mars, we need to consider the physical processes by which primary particles arriving at Mars interact with the Martian atmosphere.
Generally speaking, primary particles (Galactic Cosmic Rays (GCR) or Solar Energetic Particles (SEP)) that reach Mars may undergo elastic or inelastic nuclear interactions with atmospheric  nuclei, losing energy  and, in inelastic reactions, creating secondary particles via spallation and fragmentation processes.  
These secondary particles may further interact with the atmosphere as they propagate downwards and even with the Martian regolith, and finally result in complex spectra including both primaries and secondaries at the surface of Mars.
There are various particle transport codes such as HZETRN \citep{slaba2016solar, wilson2016}, PHITS \citep{sato2013} and GEANT4/PLANETOCOSMICS \citep{desorgher2006planetocosmics} and the newly developed GEANT4/AtRIS tool \citep{banjac2018} which can be employed for calculating the interaction of particles with the Martian atmospheric and regolith environment.

These transport models need to be implemented with a model containing Martian atmosphere and regolith properties, shown as the second row in the third column of the flowchart shown in Fig. \ref{fig:model_chart}. In particular, the Mars Climate Database \citep[MCD, http://www-mars.lmd.jussieu.fr;][]{lewis1999climate} offers location- and time-dependent descriptions of the Martian atmospheric properties, such as temperature, density, composition, etc.  
In the current model setup, we use the atmosphere profiles from MCD above the ground at the location of Gale Crater, which is the landing site of the Curiosity rover, in order to validate the modeled surface radiation field with actual RAD measurement, as discussed later.
More detailed descriptions of features of the MCD implemented in the model used here can be found in \citet{guo2018generalized, guo2019atris} and \citet{appel2018}.

Based on particle transport models implemented with the Martian environment features, the radiation on Mars can be calculated. Previous works have studied the Martian surface radiation under different conditions, e.g., at different geographic locations \citep[e.g.,][]{saganti2004radiation}, under various Martian atmospheric environment such as during dust storms \citep[e.g.,][]{norman2014influence} or at different atmospheric depth \citep{guo2017dependence} or in the Martian soil \citep{dartnell2007modelling}, with GCRs under different solar modulation conditions \citep[e.g.,][]{simonsen1992mars}, with SEPs of different spectra \citep[e.g.,][]{townsend2006carrington, guo2018generalized}, etc.
Comparison of GEANT4 and HZETRN for calculating the GCR radiation environment on Mars has shown consistent results from both models \citep{gronoff2015computation}.
However, only recently have the surface particle spectra and dose detected by RAD allowed for direct validation of such models with actual measurements (second row of the last column in Fig. \ref{fig:model_chart}).
Particularly relevant for the current study, the GEANT4 model has been shown to predict the surface particle spectra with good accuracy for many (such as hydrogen and helium isotopes and heavy ions), but not all aspects (such as neutrons), as validated by RAD measurements of the GCR-induced radiation \citep{matthia2016martian,MATTHIA201718, guo2019atris}.
The above-described model set up, implementation and validation is summarized as Step I in Fig. \ref{fig:model_chart}. To meet the requirement of the statistics of the modeling results, especially from ions heavier than protons, the computational time needed could be as large as 10 thousand CPU h or more.  

Based on the validated GEANT4 particle transport model through the Martian atmosphere, we have calculated the induced Martian surface radiation environment of more than 30 significant solar events (measured in situ at Earth), in order to provide insights into the possible variety of Martian surface radiation environments that may be induced during large SEP events \citep{guo2018generalized}.
Depending on the intensity and shape of the solar particle spectra incident at the top of the atmosphere, as well as the distribution of particle types, different SEP events lead to rather different radiation environments on the surface of Mars.
We first studied some historical SEP events and their induced radiation on Mars, i.e., deposited dose rate and dose equivalent rate in a 0.5 mm water slab accumulated from surface secondary particle spectra of various types (proton, electron, position, $^4$He, $^3$He, deuteron, triton, neutron and gamma, etc.) induced by an SEP event.
According to the International Commission on Radiological Protection (ICRP) recommendations \citep{icrp60}, the dose equivalent is the re-weighted dose by a quality factor $\rm{Q}$ which is a function of Linear Energy Transfer (LET, energy deposited per distance).  
We found that the surface dose or dose-equivalent rates do not depend significantly on the full primary spectra, which could range from a few keV up to several hundreds MeV (occasionally even reaching 1 or 2 GeV or further above).
For both the October 1989 and September 1989 events which have energetic protons up to about 2 GeV, the surface dose rate resulting from 15-1000 MeV primary protons is about 96\% of that from the full SEP spectra; the surface dose rate resulting from 100-800 MeV primary protons is about 93\% of that induced by the full SEP spectra.
This is because of the effective atmospheric shielding of particles with low energies (protons below about 150 MeV hardly penetrate down to the surface depth at $\sim$ 20 g/cm$^2$) and the nature of the SEP spectra where the high energy component has significantly less flux.
The above implementation of the validated model to calculate the SEP-induced Martian surface radiation is summarized as Step II in Fig. \ref{fig:model_chart}.

In this study, we calculate the induced Martian surface radiation by a variety of SEP events with different properties such as their energy range, intensity, power-law index, etc as described in the main text.
Then we analyze and correlate the parametrized properties of the SEP spectra and the resulting surface dose rate of each event in order to quantify the Martian surface radiation based on SEP spectral properties.
We find that at the pivot energy of $\sim 300$ MeV, the SEP intensity alone can determine the surface dose rate. 
This correlation minimizes the computational power of deriving the SEP-induced Martian surface radiation level to less than a millisecond, shown as Step III in Fig. \ref{fig:model_chart}. The result and discussion of the pivot energy study are elaborated in the main text. 

\textbf{Appendix B: Mathematical Derivation of the Pivot Energy}:
\\
Any given SEP power-law spectrum can be written as:
\begin{eqnarray}\label{eq:pivot_math1}
I(E) = I_0 E^{\gamma}.
\end{eqnarray}
For certain energy, the above function can be re-written as: 
\begin{eqnarray}\label{eq:pivot_math11}
I(E) = I_{E_i} (\frac{E}{E_i})^{\gamma}, 
\end{eqnarray}
where $I_{E_i}$ is the SEP intensity at certain energy $E_i$, e.g., the x-axes of SEP intensity at 100, 200, 300, or 500 MeV as shown in Fig. \ref{fig:pivot_point}(a)-(d). 
For a given power-law index $\gamma$, the calculated Martian surface dose rate can be fitted linearly proportional to the SEP intensity at certain energy with a function $y=c \cdot x$, as shown in Fig. \ref{fig:pivot_point}. For the same event (fixed gamma), we obtain a certain surface dose rate so that: 
\begin{eqnarray}\label{eq:pivot_math2}
 c_i I_{E_i}  = c_j I_{E_j}
\end{eqnarray}
Combining the above two equations, we obtain: 
\begin{eqnarray}\label{eq:pivot_math3}
\frac{c_i}{c_j}= (\frac{E_i}{E_j})^{-\gamma}
\end{eqnarray}
Applying the above equation with two different power-law indices $\gamma_1$ and $\gamma_2$, we have: 
\begin{eqnarray}\label{eq:pivot_math4}
c_{i1}= c_{j1} (\frac{E_i}{E_j})^{-\gamma_1} \\ \nonumber
c_{i2}= c_{j2} (\frac{E_i}{E_j})^{-\gamma_2}
\end{eqnarray}
For the existence of a pivot energy at $E_i$, we expect the correlation coefficient to be constant for different $\gamma$ values, i.e., $c_{i1}=c_{i2}$.  Implementing this relation to Eq. \ref{eq:pivot_math4}, we obtain \footnote{Note that there is a typo, which has been corrected here, in the following equation of the published version here https://doi.org/10.3847/2041-8213/ab3ec2. }:
\begin{eqnarray}\label{eq:pivot_math5}
E_p \equiv E_i = (\frac{c_{j1}}{c_{j2}})^{\frac{1}{\gamma_1-\gamma_2}} E_j
\end{eqnarray}
This means that for a given SEP energy, e.g., $E_j$, with two coefficients ${c_{j1}}$ and ${c_{j2}}$ derived under two different power-law indices we can obtain the pivot energy $E_i$, also defined as $E_p$. 
For instance, using values of $\gamma_1=-2$ and $\gamma_2=-3$ as well as ${c_{j1}}=11.02$ and ${c_{j2}}=18.44$ for SEP intensity at $E_j =500$, as shown in Fig. \ref{fig:pivot_point}(d), we can derive $E_i$ to be 299 MeV, $\sim$ 300 MeV. Such estimations base on different $E_j$ and fitted $c_j$ may differ from one another. But the average $E_p$ is around 300 MeV.

\end{appendix}

\end{document}